\documentclass[conference]{IEEEtran}
\ifCLASSINFOpdf
\else
\fi
\usepackage{epsfig,endnotes}
\usepackage{epsfig,endnotes}
\usepackage{amsfonts}
\usepackage{capt-of}
\usepackage{wrapfig}
\usepackage{wrapfig,booktabs}
\usepackage{lipsum}
\usepackage{adjustbox}
\usepackage{epsfig}
\usepackage{caption}
\usepackage{hhline}
\usepackage{amsmath}
\usepackage{xfrac}
\usepackage{graphicx}
\usepackage[font=footnotesize,caption=false]{subfig}

\usepackage{stmaryrd}
\usepackage{array}
\usepackage{mdwmath}
\usepackage{mdwtab}
\usepackage{eqparbox}
\usepackage{stfloats}
\usepackage{moreverb}
\usepackage{listings}
\usepackage{algpseudocode}
\usepackage{algorithm}
\usepackage{url}
\usepackage{color}
\usepackage{multirow}
\usepackage{setspace}
\usepackage[utf8]{inputenc}
\usepackage[T1]{fontenc}
\usepackage{microtype}
\usepackage{datetime}
\usepackage{epsfig}
\usepackage{wrapfig,lipsum,booktabs}
\usepackage{longtable}
\usepackage{xspace}
\usepackage{tabu}
\usepackage{boldline}
\usepackage{balance}

\newcommand{\aaf}{\vspace*{-6pt}}
\newcommand{\af}{\vspace*{-3pt}}

\newcommand{\todo}[1]{{\textcolor{black}{#1}}}

\begin{document}

\title{A Cross-Architecture Instruction Embedding Model for 
Natural Language Processing-Inspired Binary Code Analysis}

\author{\IEEEauthorblockN{Kimberly Redmond,  \quad Lannan Luo, \quad Qiang Zeng}
              \IEEEauthorblockA{University of South Carolina \\               redmonkm@email.sc.edu, \{lluo, zeng1\}@cse.sc.edu}}

\maketitle

\begin{abstract}

Given a closed-source program, such as most of proprietary software and  
viruses, binary code analysis is indispensable for many tasks, such 
as code plagiarism detection and malware analysis.
Today, source code is very often \emph{compiled for various architectures},
making cross-architecture binary code analysis increasingly important. 
A binary, after being disassembled, is expressed 
in an assembly languages. 
Thus, recent work starts exploring Natural Language Processing (NLP)
inspired binary code analysis.
In NLP, words are usually 
represented in high-dimensional vectors (i.e., embeddings) 
to facilitate further processing, which is one of the most
common and critical steps in many NLP tasks. 
We regard \emph{instructions as words} in NLP-inspired binary code analysis,
and aim to represent instructions as embeddings as well.

To facilitate cross-architecture binary code analysis, our
goal is that similar instructions, \emph{regardless of their architectures}, have
embeddings close to each other. To this end, we  propose a \emph{joint learning} 
approach to generating instruction embeddings that capture not only 
the semantics of instructions within an architecture, but also their \emph{semantic 
relationships across architectures}. To the best of our knowledge, this is the 
first work on building cross-architecture instruction embedding model.
As a showcase, we apply the model to resolving one of the most fundamental 
problems for binary code similarity comparison---semantics-based basic block 
comparison, and the solution outperforms the code statistics based approach. 
It demonstrates that it is promising to apply the model
to other cross-architecture binary code analysis tasks.  
\end{abstract}

\section{Introduction} \label{into}

When the source code of programs is not available, binary code analysis becomes 
indispensable for a variety of important tasks, such as plagiarism 
detection~\cite{luo2014semantics,vapd-journal}, 
malware classification~\cite{zhang2014semantics,iwamoto2012malware}, and vulnerability 
discovery~\cite{pewny2015cross,ccs2017graphembedding}.
Increasingly, software is \emph{cross-compiled for various architectures}. For example, 
hardware vendors often use the same code base to compile firmware for different devices 
that operate on varying architectures (e.g., x86 and ARM): this could cause a single 
vulnerability at source-code level to spread across binaries across diverse devices. 
As a result, \emph{cross-architecture binary code analysis} has become an emerging 
problem that draws great attention~\cite{pewny2015cross,eschweiler2016discovre,
feng2016scalable,ccs2017graphembedding, zuo2018neural}. Analysis of binaries across
instruction set architectures (ISAs), however, is non-trivial: binaries of varying 
ISAs differ greatly in instruction sets; calling conventions; 
general- and special- purpose CPU register usages; and memory addressing modes.

A binary, after being disassembled, is expressed in an assembly language. 
Given this insight, binary code analysis can be approached 
by borrowing ideas and techniques of Natural Language Processing 
(NLP)---a rich area focused on processing texts from various natural
languages~\cite{chowdhury2003natural}. In many NLP tasks, words are first often converted into 
\emph{word embeddings} (i.e., high-dimensional vectors) to facilitate further 
processing~\cite{wieting2015towards,tang2014learning}. A word's embedding is able to capture the contextual semantic 
meaning of the word; thus, words that have similar contexts have embeddings that 
appear close together in the high-dimensional space~\cite{mikolov2013efficient}. 

We regard \emph{instructions as words} in NLP-inspired binary code analysis,
and thus aim to represent instructions as embeddings as well.
To facilitate cross-architecture binary code analysis, our
goal is that similar instructions, \emph{regardless of their architectures}, have
embeddings that are close in the high dimensional space. 
Specifically, we aim to learn semantic features 
for each instruction, such that instructions \emph{in one architecture} with 
similar semantics are assigned similar vector representations (the 
\textbf{mono-architecture} objective); and additionally,  instructions 
\emph{across different architectures} with similar semantics have similar 
vector representations (the \textbf{cross-architecture} objective). 
We call such vector representations \emph{cross-architecture instruction embeddings}.

\noindent \textbf{\emph{Why Cross-Architecture Instruction Embeddings?}} 
The cross-architecture instruction embeddings capture semantic relations
of instructions across architectures, and keep invariant among tasks.
Thus, it has many potential applications to cross-architecture
binary code analysis.
Take the search of semantically equivalent functions 
as an example. Given a function in x86
that contains, for instance, the \emph{Heartbleed} function, 
by searching for functions similar to it from
a large database of functions of varying architectures,
more vulnerability instances may be found. 
This question has gained intense research
interest~\cite{pewny2015cross,eschweiler2016discovre,
feng2016scalable,ccs2017graphembedding}. 
A core subtask involved is the comparison
of basic blocks across architectures. 
We will show how the proposed technique
can be applied to resolving this subtask.

Moreover, many deep learning based NLP techniques take
word embedding as inputs. Following the idea of NLP-inspired
binary code analysis, the proposed instruction embedding model can be applied to, e.g., 
classifying binaries across architectures by feeding the instruction embeddings of the binaries 
into classic neural network structures that are used for classifying texts in NLP~\cite{lai2015recurrent,yang2016hierarchical,wang2016semantic}.

\noindent \textbf{\emph{Our Approach}.}
We propose to learn the cross-architecture instruction embedding through a 
\emph{joint learning} approach. Specifically, our joint model utilizes both 
the context concurrence information present in the instruction sequences from
the same architecture, and the semantically-equivalent signals exhibited in 
the instruction sequence pairs from different architectures. By jointly
learning these two types of information, our model can achieve both the 
\emph{mono-architecture} and \emph{cross-architecture} objectives, and generate 
high-quality cross-architecture instruction embeddings that capture
not only the semantics of instructions within an architecture, but also their \emph{semantic 
relationships across architectures}.

We have implemented the novel cross-architecture instruction embedding model, 
and conducted a series of experiments to 
evaluate the quality of the learned instruction embeddings.
Moreover, as a showcase, we apply the model to resolving one of
the most fundamental problems for binary code similarity comparison, that is,
semantics-based basic block similarity comparison. Our solution achieves 
AUC = 0.90. Recent work~\cite{eschweiler2016discovre,feng2016scalable,ccs2017graphembedding} 
uses several manually selected statistic features (such as the number of instructions 
and the number of constants) of a basic block to represent it. However, a SVM classifier based on such features only achieves AUC = 0.85 for the same task.
The trained models, datasets, and evaluation results are publicly 
available.\footnote{https://github.com/nlp-code-analysis/cross-arch-instr-model}

We summarize our contributions as follows:

\af
\begin{itemize}

\item 
To the best of our knowledge, this is the \emph{first} work on building a uniform
cross-architecture instruction embedding model that tolerates the
significant syntactic differences across architectures. 

\item 
We propose an effective joint learning approach to training
the model, which makes use of both the information
in the instruction sequences from the same architecture, 
and the semantically-equivalent signals exhibited in the instruction 
sequence pairs from different architectures.

\item 
We implement model, and the evaluation demonstrates the good 
quality of the learned instruction embeddings.
Moreover, we apply the model to cross-architecture
basic block similarity comparison and the solution
outperforms the statistic feature based approach.

\item
This research successfully demonstrates that it is promising %and possible 
to adapt NLP ideas and techniques to binary code analysis tasks. 
Just like word embeddings are critical for many NLP tasks,
the proposed instruction embedding model can substantially facilitate
NLP-inspired cross-architecture binary code analysis.

\end{itemize}

\section{Related Work}
We expect the proposed model can be naturally applied to
binary code similarity comparison. Existing binary code analysis techniques for code
similarity comparison can be roughly divided 
into two classes: traditional approaches and machine learning-based ones. 

\noindent \textbf{Traditional approaches.}
Most traditional approaches work on a \emph{single} architecture. First, static 
plagiarism detection or clone detection includes
string-based~\cite{baker1995finding,bilenko2003adaptive,deerwester1990indexing}, 
AST-based~\cite{jiang2007deckard,truong2004static,yang1991identifying,
koschke2006clone},
token-based~\cite{kamiya2002ccfinder, schleimer2003winnowing, prechelt2000jplag}, 
and PDG-based~\cite{gabel2008scalable, liu2006gplag,crussell2012attack,li2012cbcd}.
\emph{Source code-based} approaches are inapplicable for closed-source software.
Symbolic execution of binary code has been enabled by tools such as 
BitBlaze~\cite{bitblaze} and BAP~\cite{bap}; it is very accurate in extracting 
code semantics~\cite{luo2017semantics}, but it is very 
\emph{computationally expensive} and \emph{unscalable}. % to a large codebase. 

Recent works have applied traditional approaches to addressing the 
\emph{cross-architecture} scenario~\cite{pewny2015cross,eschweiler2016discovre,
chandramohan2016bingo,feng2017extracting,david2017similarity,david2018firmup,
david2016statistical}. Multi-MH and Multi-k-MH~\cite{pewny2015cross} are the 
first two methods for comparing functions of different ISAs. But their 
fuzzing-based basic-block similarity comparison and graph (i.e., CFG) 
matching-based algorithms are very expensive. 
discovRE~\cite{eschweiler2016discovre} boosts CFG-based matching process, 
but is still expensive.
Both \texttt{Esh}~\cite{david2016statistical} and its
successor~\cite{david2017similarity} use data-flow 
slices of basic blocks as the basic comparable unit. 
\texttt{Esh} uses SMT solver to verify function similarity, which makes it 
unscalable. In \cite{david2017similarity}, binaries are lifted to IR for 
creating function-centric signatures.

\noindent \textbf{Machine learning based approaches.}
Machine learning, including deep learning, has been applied to code analysis~\cite{chua2017neural,lee2017learning,feng2016scalable,ccs2017graphembedding,
mokhov2014use,luo2016solminer,mou2016convolutional,huo2016learning,white2017sorting,
huo2017enhancing,nguyen2017exploring,han2017learning,dingasm2vec}. 
Lee et al.\ propose \texttt{Instruction2vec} for converting assembly instructions 
to vector representations~\cite{lee2017learning}; but their instruction 
embedding model can only work on a \emph{single} architecture. 
\texttt{Asm2Vec}~\cite{dingasm2vec} produces a numeric vector for each function, 
but can only work on a \emph{single} architecture.
We instead build 
a \emph{cross-architecture instruction embedding model} which works
for varying architectures.

A few works target \emph{cross-architecture} binary code 
analysis~\cite{feng2016scalable,ccs2017graphembedding,chandramohan2016bingo,
zuo2018neural}. Some exploit the \emph{statistical} aspects 
of code, rather than its semantics. For example, Genius~\cite{feng2016scalable} 
and Gemini~\cite{ccs2017graphembedding} use some \emph{manually} selected 
statistical features (e.g., the number of constants) to represent basic 
blocks, but they ignore \emph{the meaning of instructions and the 
dependency between them}, resulting in significant loss of semantic 
information. \texttt{INNEREYE-BB}~\cite{zuo2018neural} uses LSTM
to encode each basic block into an embedding, but it needs to train 
a \emph{separate} instruction embedding model for each architecture. 
Instead, we build a uniform cross-architecture instruction embedding 
model that tolerates the syntactic differences across architectures.

\noindent \textbf{Summary.} 
To the best of our knowledge, ours is the \emph{first work} 
to learn cross-architecture instruction embeddings that capture 
semantic features \emph{invariant 
to specific tasks}. Such instruction embeddings can be 
adopted to a variety of important code analysis tasks, and help us scale 
to more architectures.

\section{Background}

\subsection{Word Embeddings}

% ~\cite{mikolov2013efficient, mikolov2013distributed,mnih2013learning,dhillon2011multi}

Many NLP models applying deep learning techniques have been proposed to learn 
high-quality word embeddings, with Mikolov's \emph{skip-gram} (SG) and 
\emph{Continuous Bag Of Words} (CBOW)~\cite{mikolov2013efficient} gaining a 
lot of traction due to their relatively low memory use, and overall 
increased efficiency.

%The skip-gram model learns word embeddings using a neural network. During training, 
%a sliding window is employed on a text stream. In Figure~\ref{fig:skip-gram}, for
%example, a window of size 2 is used, covering two words behind the current word 
%and two words ahead. The model starts with a random vector for each word, and then 
%gets trained when going over each sliding window. In each sliding window, the embedding 
%of the current word, \textbf{w}$_{t}$, is used as the parameter vector of a \emph{softmax} 
%function (Equation~\ref{equ_word}) that takes an arbitrary word $w_k$ as a training 
%input and is trained to predict a probability of 1, if $w_k$ appears in the context 
%$C_t$ (i.e., the sliding window) of $w_t$, and 0, otherwise. 

%the objective is to predict a word’s context given the word itself. 

The SG model takes each word as the input and predicts the context corresponding to
the word, while the CBOW model takes the context of each word as the input and 
predicts the word corresponding to the context. During training, a sliding window 
is applied on a text. Each model starts with a 
random vector for each word, and then gets trained when going over each sliding window. 
After the model is trained, the embeddings of each word become meaningful, yielding 
similar vectors for similar words. Due to their simplicity, 
% and the use of the hierarchical softmax, 
both models can achieve very good performances on various semantic tasks, 
and can be trained on a desktop computer at billions of words per hour. 
%Plus, learning the word embedding is entirely \emph{unsupervised}.

%After the model is trained on many sliding windows, the embeddings of each word 
%become meaningful, yielding similar vectors for similar words. 
%The \emph{skip-gram} model has a very simple neural network architecture,
%containing only a linear projection input layer and an output softmax layer, 
%Despite its simplicity, the \emph{skip-gram} model can achieve very good 
%performances on various semantic tasks while having an 
%advantage of fast training time.

\subsection{Multilingual Word Embeddings}

%Recent work in learning multilingual representations tend to tailor towards 
%achieving good performance on multilingual tasks, most often the crosslingual 
%document classification tasks.

A wide variety of multilingual NLP tasks, including machine 
translation~\cite{brown1993mathematics}, entity clustering~\cite{green2012entity}, 
and multilingual document classification~\cite{bel2003cross}, have motivated 
recent work in training \emph{multilingual word representations} where similar-meaning 
words in different human languages are embedded close together in the same 
high-dimensional space.

Different approaches have been proposed to training multilingual word embeddings.
For example, one category of approaches is based on \emph{multilingual mapping},
where word embeddings are first trained on each language independently and 
a mapping is then learned to transform word embeddings from one language 
to another~\cite{mikolov2013exploiting}. Another category
\todo{attempts} to \emph{jointly learn} multilingual word embeddings from scratch~\cite{klementiev2012inducing,kovcisky2014learning,
gouws2015bilbowa,luong2015bilingual}. 
Our cross-architecture 
instruction embedding model adapts the technique proposed in~\cite{luong2015bilingual}.

%\section{System Design}
\section{Cross-Architecture Instruction \\ Embedding Model}
In light of the idea of \todo{NLP-inspired} binary code analysis,
we regard \emph{instructions as words}.
An instruction includes an opcode (specifying the instruction operation) 
and zero or more operands (specifying registers, memory locations, or literal 
data). For example, \texttt{mov ebp, esp} is an instruction where \texttt{mov} 
is an opcode and both \texttt{ebp} and \texttt{esp} are operands.
Note that the assembly code in this paper adopts the Intel syntax.

\subsection{Design Goal}

Our goal \todo{in} building the instructions model is to achieve both the \emph{mono-architecture}  and 
\emph{cross-architecture} objectives. That is, we want the learned 
cross-architecture instruction embeddings not only \todo{to} preserve the clustering 
properties mono-architecturally (instructions in one architecture with 
similar semantics are close together in the vector space), but also 
exhibit the semantic relationships across 
different architectures (instructions across architectures with 
similar semantics are close together).

\subsection{System Overview}

Our proposed cross-architecture instruction embedding model adapts the 
\emph{joint learning} approach in~\cite{luong2015bilingual}, consisting 
of a \emph{mono-architecture} component and a \todo{\emph{multi-architecture}} 
component. The \emph{mono-architecture} component utilizes the context 
concurrence information present in the input instruction sequences from 
the same architecture  (Section~\ref{sec:mono-comp}); and the 
\emph{multi-architecture} component learns the semantically-equivalent 
signals exhibited in the equivalent instruction sequence pairs from varying 
architectures  (Section~\ref{sec:multi-comp}).

An \emph{instruction sequence} in our work is a basic block, as
we regard instructions as words and basic blocks as sentences. 
Note that we do not consider a function as a sentence, as a function cannot 
be treated as a straight-line sequence: when a function is invoked, its instructions
are not executed sequentially.

\noindent \textbf{Handling {out-of-vocabulary} (OOV) instructions.}
The issue of OOV words is a well-known problem in NLP,
and it exacerbates significantly in our case as constants, address offsets, labels, 
and strings are frequently used in instructions. To address it, instructions 
are preprocessed using the following rules: 
(1) \todo{Numerical} constant values are replaced with 0, and the minus signs are preserved.
(2) \todo{String} literals are replaced with \verb|<STR>|.
(3) \todo{Function} names are replaced with \verb|FOO|.
(4) Other symbol constants are replaced with \verb|<TAG>|.

\noindent \textbf{Joint objective function.}
Below is our \emph{joint objective} function:

\aaf\aaf
\begin{equation} \label{all-obj}
J = \gamma \sum_{i=1}^{N} J(\texttt{Mono}_{a_i}) + \beta \sum_{i=1}^{N-1} \sum_{j=i+1}^{N} J(\texttt{Multi}_{<a_i, a_j>})
\end{equation}

In this equation, each mono-architecture component, \texttt{Mono}$_{a_i}$
($\forall i \in \{1, ... N\}$) aims to capture the clustering property of 
the corresponding architecture $a_i$, where $J(\texttt{Mono}_{a_i})$ is the 
objective function of \texttt{Mono}$_{a_i}$. Each \todo{multi-architecture} component, 
\texttt{Multi}$_{<a_i, a_j>}$ ($\forall i, j \in \{1, ... N\}$, $i \neq j$), 
is used to learn the semantic relationships across architectures,
where $J(\texttt{Multi}_{<a_i, a_j>})$ is the 
objective function of \texttt{Multi}$_{<a_i, a_j>}$.
The $\gamma$ and $\beta$ hyperparameters balance out the influence of the 
mono-architecture components over the \todo{multi-architecture} one. 

Specifically, if there are only two architectures, e.g., x86 and ARM, the 
\emph{joint objective} function becomes:

\aaf\aaf
\begin{equation} \label{two-obj}
J = \gamma (J(\texttt{Mono}_{x86}) + J(\texttt{Mono}_{ARM})) + \beta  J(\texttt{Multi}_{<x86, ARM>})
\end{equation}

\subsection{Mono-Architecture Component} \label{sec:mono-comp}

Any word embedding model can be a candidate to be selected to build the
mono-architecture component~\cite{mikolov2013efficient,klementiev2012inducing,
gouws2015bilbowa, mikolov2013distributed,mnih2013learning,dhillon2011multi}. 
Based on our experiment, we adopt the CBOW model as implemented in 
\texttt{word2vec}~\cite{mikolov2013efficient}, which achieves better performance
than the skip-gram model.

The CBOW model predicts a current instruction based on its context. 
During training, a sliding window with size $n$ is employed on an instruction 
sequence. \todo{The} context of a current instruction $e_t$ is defined as $n$ instructions 
before and after $e_t$ within the corresponding sliding window.
The CBOW model contains three layers. The input layer corresponds to the 
context. The hidden layer corresponds to the projection of each instruction from 
the input layer into the weight matrix, which is then projected into the third 
output layer. The final step is the comparison between the output and the current 
instruction in order to correct its vector representation based on the back propagation 
of the error gradient. Thus, the \emph{objective} of the CBOW model is to 
maximize the following equation:

\aaf\aaf
\begin{equation} \label{equ_window}
J = \frac{1}{T} \sum_{t=1}^{T} \texttt{log} \hspace{2pt} P(e_t|e_{t-n},...e_{t-1},e_{t+1},...,e_{t+n})
\end{equation} 
where $T$ is the length of the instruction sequence, and $n$ the size of the 
sliding window.

After the model is trained on many sliding windows, similar instructions 
tend to have embeddings that appear close together in the \todo{high-dimensional} 
vector space.

\subsection{Multi-Architecture Component} \label{sec:multi-comp}

\begin{figure} 
\centering
\includegraphics[scale=0.68]{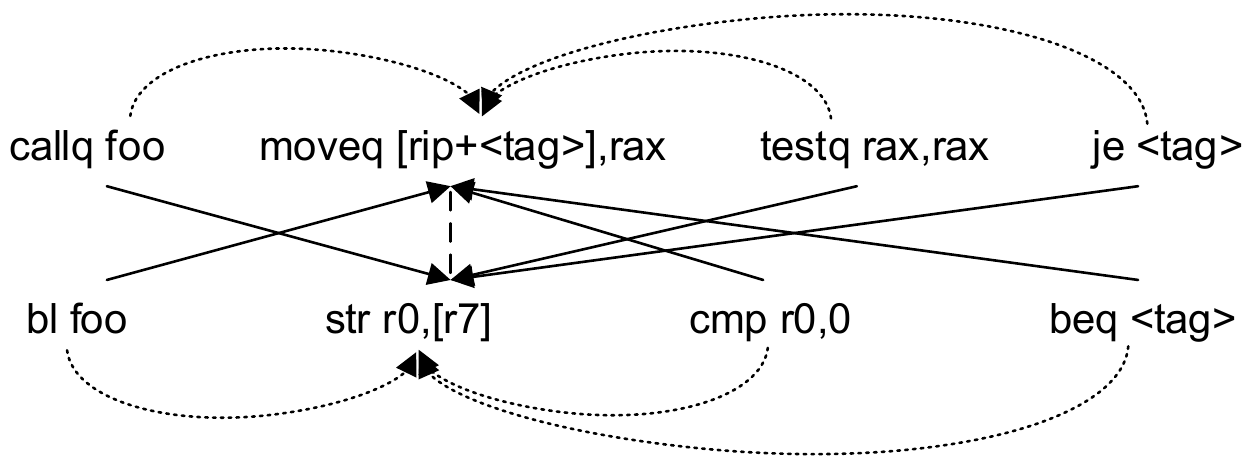}
%\captionsetup{font={small}}
\caption{A cross-architecture instruction embedding model. 
}\label{fig:multi}
\aaf\aaf
\end{figure}

We adopt the CBOW model to build our \todo{multi-architecture} component. 
Our cross-architecture instruction embedding model is extended from 
the CBOW model as implemented in \texttt{word2vec}, and is 
effective in learning instruction representations both mono-architecturally 
and multi-architecturally.

Figure~\ref{fig:multi} shows how our cross-architecture instruction embedding
model works. The input is a pair of semantically-equivalent basic blocks, 
each of which is a sequence of instructions: the instruction sequence of the 
basic block compiled for x86 is \{\texttt{callq foo}; 
\texttt{moveq [rip+<tag>],rax}; \texttt{testq rax,rax}; \texttt{je <tag>}\}, 
and the instruction sequence of the block compiled for ARM is 
\{\texttt{bl foo}; \texttt{str r0,[r7]}; \texttt{cmp r0,0}; 
\texttt{beq <tag>}\}. Note that the instruction sequences may be of different 
lengths, and the lengths can vary from example to example; both can be easily 
handled by the model. 

To predict instructions cross-architecturally rather than \emph{only} 
mono-architecturally as in the standard CBOW model 
(Section~\ref{sec:mono-comp}), we use the contexts in one architecture 
to predict the instructions in another architecture. 
For example, if we know that the instruction \texttt{moveq [rip+<tag>],rax}
in x86 has the same meaning as the instruction \texttt{str r0,[r7]} in
ARM, we can simply substitute \texttt{str r0,[r7]}  and use the surrounding 
instructions---such as \texttt{bl foo} and \texttt{cmp r0,0}---to predict 
\texttt{moveq [rip+<tag>],rax}.
Therefore, given an alignment link between an instruction $e_1$ in 
\todo{one} architecture and an instruction $e_2$ in another architecture, 
our cross-architecture model uses the neighbors of the instruction $e_2$ 
to predict the instruction $e_1$, and vice \todo{versa}. Then, 
$J(\texttt{Multi}_{<a_i,a_j>})$ in Equation~\ref{all-obj}, where $a_i$ 
and $a_j$ represent two different architectures, is also the \emph{objective} 
function in Equation~\ref{equ_window}. 

The challenge here is how to find the alignment links between 
instructions. There are two solutions.
(1) A simple way is to assume linear alignments between instructions across 
architectures. That is, each instruction in one sequence $M$ at position $i$ 
is aligned to the instruction in another sequence $N$ at position 
$i \times |N|/|M|$, where $|M|$ and $|N|$ are the length of the 
corresponding sequences. 
(2) Another way is to determine the alignment links based on the \emph{opcode} 
contained in each instruction. For example, from the opcode references of 
x86~\cite{x86-opcode} and ARM~\cite{arm-opcode}, we can find that \texttt{moveq} 
from x86 and \texttt{str} from ARM can be used to store data in registers;
thus, it is reasonable to align an \todo{instruction} containing \texttt{moveq} with 
another instruction containing \texttt{str}. 
Then, a dynamic programming algorithm similar to the solution to finding the Longest Common Subsequence 
can be used to determine the best alignment between two sequences.

We adopt the first solution in our current implementation. Our preliminary
results show that the model has good performance.
We plan to explore the second solution to further improve the model 
as future work.

\section{Evaluation}

This section presents our evaluation results. 
We first describe the dataset used in our evaluation (Section~\ref{sec:dataset}) 
and discuss how the model is trained (Section~\ref{sec:train-model}). We then 
conduct three different tasks to evaluate the quality of the learned model: (1) the mono-architecture instruction similarity task
(Section~\ref{sec:mono-eval}); (2) the cross-architecture instruction similarity
task (Section~\ref{sec:cross-eval}); and (3) as a concrete application, the cross-architecture basic-block 
similarity comparison task (Section~\ref{sec:bb-eval}).

\subsection{Dataset} \label{sec:dataset}

We train our model using basic blocks that are open-sourced by our prior 
work\footnote{https://nmt4binaries.github.io}~\cite{zuo2018neural}, consisting 
of 202,252 semantically similar basic-block pairs. This dataset is prepared using 
\texttt{OpenSSL} (v1.1.1-pre1) and four popular Linux packages, including 
\texttt{coreutils} (v8.29), \texttt{findutils} (v4.6.0), \texttt{diffutils} 
(v3.6), and \texttt{binutils} (v2.30). Each program is compiled by two 
architectures (x86-64 and ARM) and \texttt{clang} (v6.0.0) with three different 
compiler optimization levels (O1-O3).

Two basic blocks of different ISAs compiled from the same piece of source code
are considered as equivalent. To collect such ground truth, we modify the 
\emph{backends} of various architectures in the LLVM compiler to add the 
basic-block boundary annotator, which annotates a unique ID for each 
block so that \emph{all blocks compiled from the same piece of source code, 
regardless of architecture, will obtain the same ID}.

\subsection{Model Training} \label{sec:train-model} 

We use the following settings to train our cross-architecture instruction 
model\footnote{See~\cite{mikolov2013efficient} for more details on the 
parameters.}: the instruction embedding dimension of 200, the 
sliding window size of 5, a subsampling rate of 1$e$-5, 
negative sampling with 30 samples, and the learning rate of 0.05.
The model is trained for 10 epochs and the learning rate is decayed to 
0 once training is done. We set the hyperparameters in Equation~\ref{all-obj} 
to 1 for $\gamma$ and 4 for $\beta$. % in our experiments.

\begin{figure*}%[!htb]
    \centering
    \begin{minipage}{.32\textwidth}
	\includegraphics[height=4.9cm]{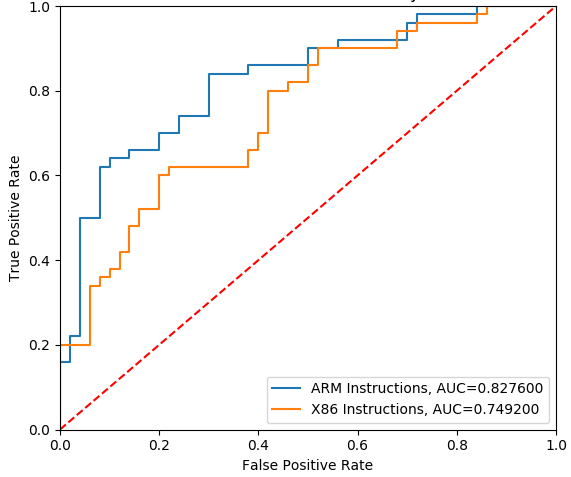}
	\caption{The ROC for the mono-architecture instruction similarity test.} 
	\label{fig:mono-instr-similarity}
    \end{minipage}%
\quad
    \begin{minipage}{.32\textwidth}
	\centering
	\includegraphics[height=4.9cm]{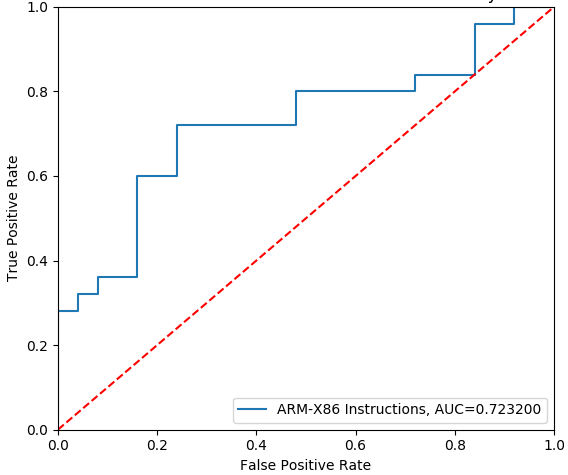}%scale=0.4 raise=-6.8pc
	\caption{The ROC for the cross-architecture instruction similarity test.}
	\label{fig:cross-instr-similarity}
    \end{minipage}%
\quad
    \begin{minipage}{.32\textwidth}
	\centering
	\includegraphics[height=4.9cm]{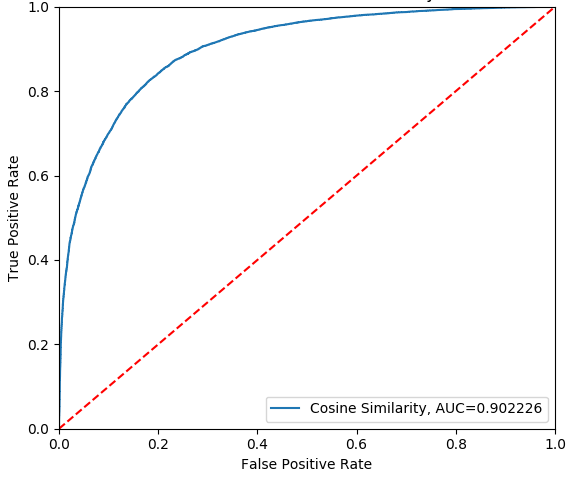}%scale=0.4 raise=-6.8pc
	\caption{The ROC for the cross-architecture basic-block similarity comparison test.}
	\label{fig:bb-similarity}
    \end{minipage}%
\aaf
\end{figure*}

\subsection{Mono-Architecture Instruction Similarity Task} \label{sec:mono-eval}

\noindent \textbf{Instruction Similarity Test.}
Unlike the case of word embedding models---which have many existing word-aligned 
corpora to evaluate the quality of word embeddings---we do not have such 
data. We thus create a set of manually-labeled instruction pairs from the same
architecture to test the instruction embeddings. 
We consider a pair of instructions to be similar if they contain the same opcode,
and a pair of instructions with different opcodes to be dissimilar; but 
a few exceptions exist---for example, in x86, \texttt{cmp} and \texttt{test} are 
different opcodes but are semantically similar, and thus instructions containing 
them tend to have similar embeddings. 
We randomly select 50 similar instruction pairs and 50 dissimilar ones, which are 
assigned with labels 1 and -1, respectively. % from the dataset. 

We then measure the similarity of two instructions based on their cosine similarity.
Figure~\ref{fig:mono-instr-similarity} shows the ROC curves and AUC values for ARM
and x86: the AUC for ARM and x86 are around 0.828 and 0.749, respectively.
It is worth noting that the accuracy of the monolingual word similarity test  
for the \emph{multilingual} word embedding models are around 0.51~\cite{dufter2018stronger}.

\begin{table*}
%\small
\caption{Nearest neighbor instructions \emph{mono-architecturally}. It shows the top two similar ARM 
instructions for four randomly selected ARM instructions as measured by cosine similarity.}
\label{tab:nearest-words}
\hspace{-10pt}
\renewcommand{\arraystretch}{1.2}
\scalebox{0.91}{
\setlength{\belowcaptionskip}{10pt}
\centering
\begin{tabular}{|c|c|c|c|c|c|c|c|}
\hline
\multicolumn{2}{|c|}{\texttt{ADD r1,r0,r7}} & \multicolumn{2}{c|}{\texttt{SUB sp,sp,0}}  & \multicolumn{2}{c|}{\texttt{LDR r0,[r5+0]}} &
\multicolumn{2}{c|}{\texttt{MOV r0,r5}}  \\
\hline
\emph{ARM}   & \emph{Sim. score}  & \emph{ARM}  & \emph{Sim. score}  & \emph{ARM}  & \emph{Sim. score}  & \emph{ARM}  & \emph{Sim. score}  \\ \hhline{|=|=|=|=|=|=|=|=|}
\texttt{ADD r1,r0,r5} & 0.643148 & \texttt{ADD r4,sp,0}   & 0.339248 & \texttt{LDR r4,[r2+0]} & 0.441962 & \texttt{LDR r2,[sp+0]} &  0.590213  \\ \hline
\texttt{ADD r1,r0,r6} & 0.630020 & \texttt{STR r4,[r0],0} & 0.339097 & \texttt{STR r7,[sp+0]} & 0.392250 &  \texttt{LDR r3,<tag>} &  0.528087  \\ \hline
\end{tabular}
}
\end{table*}

\begin{table*}
\centering
\caption{Nearest neighbor instructions \emph{cross-architecturally}. It shows the 
top two similar x86 instructions for six randomly selected ARM instructions as measured by cosine similarity.}
\label{tab:nearest-words-cross}
\hspace{-10pt}
\renewcommand{\arraystretch}{1.2}
\scalebox{1}{
\setlength{\belowcaptionskip}{10pt}
\centering
\begin{tabular}{|c|c|c|c|c|c|}
\hline
\multicolumn{2}{|c|}{\texttt{LDR r0,[r5+0]}} & \multicolumn{2}{c|}{\texttt{LDRNE r4,[sp+0]}}  & \multicolumn{2}{c|}{\texttt{ADD r1,r0,r7}}   \\
\hline
\emph{x86}   & \emph{Sim. score}  & \emph{x86}  & \emph{Sim. score}  & \emph{x86}  & \emph{Sim. score}  \\ \hhline{|=|=|=|=|=|=|}
\texttt{MOVL [rbp],eax} & 0.437663 & \texttt{CMOVLEL r14d,eax}   & 0.584823 & \texttt{MOVQ [rax+0],r13} & 0.538247  \\ \hline
\texttt{MOVL [r14+0],eax} & 0.432104 & \texttt{CMOVEQ r12,r9} & 0.584176 & \texttt{ADDQ r13,rbx} & 0.502640 \\

\hhline{======}
\hline
%\shhline[1pt]
\hhline{======}

\multicolumn{2}{|c|}{\texttt{BLT <tag>}} & \multicolumn{2}{c|}{\texttt{BEQ <tag>}}  & \multicolumn{2}{c|}{\texttt{MOV r8,r2}}  \\
\hline
\emph{x86}   & \emph{Sim. score}  & \emph{x86}  & \emph{Sim. score}  & \emph{x86}  & \emph{Sim. score}  \\ \hhline{|=|=|=|=|=|=|}
\texttt{JL <tag>} & 0.524643 & \texttt{JE <tag>} & 0.372829 & \texttt{MOVQ r13,rdi} & 0.453570  \\ \hline
\texttt{JLE <tag>} & 0.453281 & \texttt{CMPB [rsi+0],0} & 0.367084 & \texttt{MOVQ r8,rbp} & 0.413255  \\ \hline

\end{tabular}
}
\end{table*}

\noindent \textbf{Nearest Neighbor Instructions.}
We then randomly select four ARM instructions, and search for the top two 
similar ARM instructions using cosine similarity. The result is shown 
in Table~\ref{tab:nearest-words}.
We omit the result for x86 due to space \todo{limits}.
It can be observed that the learned cross-architecture instruction 
embeddings still preserve the clustering property \emph{mono-architecturally}.
For example, our embeddings find very relevant neighbor instructions
for the instruction \texttt{ADD r1,r0,r7}, such as \texttt{ADD r1,r0,r5} and \texttt{ADD r1,r0,r6}.

\subsection{Cross-Architecture Instruction Similarity Task} \label{sec:cross-eval}

\noindent \textbf{Instruction Similarity Test.}
Similar to the previous experiment, we create a set of manually-labeled cross-architecture 
instruction pairs. We first determine the similar and dissimilar \emph{opcode} pairs for 
different ISAs based on our prior knowledge and experience, and then select a set of similar 
and dissimilar \emph{instruction} pairs based on whether their contained opcodes are (dis)similar
or not. We then measure the similarity of two instructions 
using cosine similarity. Figure~\ref{fig:cross-instr-similarity} shows the ROC result. 
Our model achieves AUC = 0.723 on this test.

\noindent \textbf{Nearest Neighbor Instructions.}
We next randomly select six ARM instructions, and search for the top two 
similar x86 instructions for each ARM instruction based on their cosine similarity. 
The result is shown in 
Table~\ref{tab:nearest-words-cross}. We can see that similar instructions of 
different ISAs have embeddings close to each other, as predicted. For example, 
our embeddings find very relevant neighbor \emph{x86} instructions for the 
\emph{ARM} instruction \texttt{LDR r0,[r5+0]}, such as \texttt{MOVL [rbp],eax} 
and \texttt{MOVL [r14+0],eax}; both \texttt{LDR} from ARM and \texttt{MOV} from 
x86 can be used to load register from memory. Thus, the cross-architecture 
instruction embeddings successfully capture semantics of instructions across 
architectures.

\begin{figure}%{.49\textwidth}
	\centering
	\includegraphics[scale=0.5]{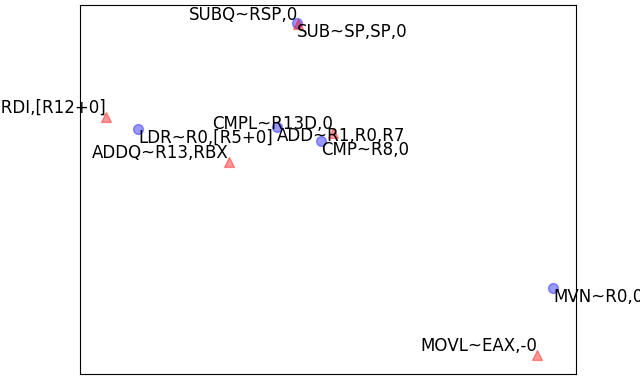}% 	
	%\captionsetup{font={small}}
	\caption{Visualization of five ARM and x86 instruction pairs. 
	A blue circle and red triangle represent an ARM and x86 instruction, 
	respectively.}\label{fig:pair-visu}
	\aaf\aaf
\end{figure}%

\noindent \textbf{Cross-Architecture Instruction Embedding Visualization.}
We next use t-SNE~\cite{maaten2008visualizing}, a useful tool for visualizing 
high-dimensional vectors, to plot all the cross-architecture instruction 
embeddings in a two-dimensional space (due to space \todo{limits}, we omit it here). 
A quick inspection shows that the instructions of different ISAs overlap together.
Our prior work~\cite{zuo2018neural} learns the \emph{mono-architecture} 
instruction embeddings, where 
the instruction embeddings are architecture-specific (i.e., separate instruction 
embedding models need to be trained for each architecture), and the embeddings 
of different ISAs exist in different vector spaces (see Figure 9 
in~\cite{zuo2018neural}). Instead, this work establishes a 
cross-architecture instruction embedding model, which learns embeddings 
in the same vector space.

We then visualize the embeddings of a set of similar instruction pairs. To this 
end, we randomly pick five x86 \todo{instructions;} and for each x86 instruction, 
we select its similar counterpart from ARM based on our prior knowledge.
We use t-SNE to plot their embeddings, as shown in Figure~\ref{fig:pair-visu}.
It can be observed that most x86 and ARM instructions with similar meanings 
appear nearby: for example, the two similar x86 and ARM instructions, 
\texttt{SUBQ RSP,0} and \texttt{SUB SP,SP,0}, are close together in the vector space. 

Therefore, our cross-architecture instruction embeddings capture not only 
instruction semantics, but also semantic relationships across different
architectures.

\subsection{Cross-Architecture Basic-Block Similarity Comparison Task} \label{sec:bb-eval}

We next conduct the cross-architecture basic-block similarity comparison task
to evaluate the quality of the learned model. 
We divide the dataset which contains 202,252 \emph{similar} basic-block pairs 
into two parts: 90\% of them are used for training; 10\% of them and another 
20,000 \emph{dissimilar} block pairs (selected from the dataset open-sourced 
by our prior work~\cite{zuo2018neural}) for testing. 
Note that we only need similar block pairs for training; and for testing both 
similar and dissimilar pairs are used. 

To measure the similarity of two basic blocks, we first compose all the instruction 
embeddings for each basic block, and then use the cosine similarity of the two
composed embeddings to measure the basic block similarity. For simplicity, we use
the sum of all the instruction embeddings of a basic block to represent it.
This simple summation has proven to be a successful way of obtaining sentence 
or document embeddings that can be used as features in specific tasks~\cite{yu2014deep, gershman2015phrase} such 
as answer sentence selection~\cite{yu2014deep}.

Figure~\ref{fig:bb-similarity} shows the ROC curve evaluated on the testing 
dataset; and our model achieves AUC = 0.90.
Recent work~\cite{eschweiler2016discovre,feng2016scalable,ccs2017graphembedding} 
looks at the statistical information of a basic block, and uses several manually 
selected features (such as the number of instructions and constants) of a basic 
block to represent it. But such an approach causes significant loss 
of information about the instructions being used and their dependencies. As a 
result, the statistics-based representation is efficient but inaccurate---a SVM 
classifier based on such features can only achieve AUC = 0.85 according to our 
prior work~\cite{zuo2018neural}.

Therefore, our model, capturing the meaning of instructions and the dependency 
between them, can provide more precise basic-block representation and efficient 
comparison.
It is worth mentioning that many prior 
systems~\cite{gao2008binhunt,luo2014semantics,
pewny2015cross,feng2016scalable,ccs2017graphembedding} built on basic-block comparison can 
benefit from our model.

\section{Future Work}

\noindent \textbf{Improvements.}
Currently, heuristics are used to decide parameter values; e.g., the window 
size is set as 20. We will investigate the stability of the cross-architecture 
instruction embedding model with respect to different hyperparameters---including 
the sliding window size, the number of epochs, and the instruction embedding 
dimension.

Two solutions are proposed in Section~\ref{sec:multi-comp} to find the alignment 
links between instructions. We have tried the first simple solution, and plan 
to explore the second one to attest how important alignment information is in 
learning cross-architecture instruction embeddings.

The sliding window based on program paths can reflect the context information 
of instructions more precisely and may generate better instruction embeddings. 
We plan to explore dynamic analysis to generate a set of semantically-equivalent
paths from two programs compiled for different architectures, and use them 
for training. We will evaluate the model trained on paths in terms of accuracy 
and efficiency.

\noindent \textbf{Applications.}
A prominent application of cross-architecture instruction embeddings 
(similar to multilingual word embeddings) is that the induced instruction 
embeddings enable us to \emph{transfer} a classifier trained on one architecture 
to another without any adaptation. We plan to investigate the transferability
by applying our model to the cross-architecture program/function classification 
problem. For example, we will train a classifier using the code compiled for x86, 
and check whether it can directly work on ARM. 

Moreover, we plan to apply our model to other important code analysis tasks, 
such as cross-architecture bug search, and compare our model to recent 
approaches~\cite{feng2016scalable,ccs2017graphembedding,zuo2018neural,david2018firmup}.

\section{Conclusion}

To the best of our knowledge, this is the first work that aims to learn 
cross-architecture instruction embeddings that tolerate the syntactic
differences of instructions across architectures and capture their important 
semantic features.
We adopt a joint learning approach to building the instruction embedding model, such that 
instructions with similar semantics, regardless of their architectures, have embeddings 
close together in the vector space. Our instruction similarity tests and 
cross-architecture basic-block similarity comparison task demonstrate the 
good quality of the learned instruction embeddings. 
The proposed  model may 
be applied to many cross-architecture binary code analysis tasks,
such as vulnerability finding, malware detection, and plagiarism detection.

\balance

\bibliographystyle{IEEEtranS} 
{
\bibliography{code-embedding.bib} 
}

\end{document}